\documentclass[aps,prc,onecolumn,nofootinbib,superscriptaddress]{revtex4-2}

\usepackage{amsmath,amssymb,bm}
\usepackage{graphicx}
\usepackage{hyperref}
\usepackage{url}

\hypersetup{
  colorlinks=true,
  linkcolor=blue,
  citecolor=blue,
  urlcolor=blue
}

\newcommand{\dd}{\mathrm{d}}
\newcommand{\ee}{\mathrm{e}}

\begin{document}

\title{Emergent equilibrium-like yields from nonequilibrium cascade dynamics}

\author{Takeshi Fukuyama}
\affiliation{Research Center for Nuclear Physics (RCNP), Osaka University, Ibaraki, Osaka 567-0047, Japan}

\begin{abstract}
We study nonequilibrium cascades in which fragile bound or coherent
structures are formed through intermediate states rather than by direct
equilibration.
Motivated by light-nuclei production in relativistic heavy-ion
collisions and by Bose--Einstein condensation in cosmological settings,
we analyze such processes within the Schwinger--Keldysh real-time
formalism.
We show that commonly used rate equations can be understood as a
controlled Markovian approximation obtained by integrating out
intermediate reservoirs in an underlying multi-component
nonequilibrium dynamics.
When the finite lifetime of these reservoirs is retained,
non-Markovian memory effects naturally appear, leading to delayed and
history-dependent formation dynamics.
The associated memory time provides a quantitative criterion
for the validity of reduced, rate-based descriptions far from
equilibrium.

Applying this framework to light-nuclei formation mediated by
$\Delta$ resonances and to condensate formation with localized
intermediate structures, we show that equilibrium-like final yields
arise as a consequence of finite-lifetime intermediate reservoirs
and memory effects, rather than from genuine thermal equilibration.
\end{abstract}

\maketitle

\section{Introduction}

Nonequilibrium dynamics governs a wide range of physical systems, from relativistic heavy-ion collisions to cosmological phase transitions.
In many such systems, weakly bound or coherent final states appear to exhibit thermal or equilibrium-like properties, despite forming in environments that are hot, rapidly evolving, and far from equilibrium.
Understanding how such states emerge and survive has become a central conceptual challenge.

Recent analyses of light-nuclei production in heavy-ion collisions have shown that deuterons and other loosely bound nuclei do not behave as elementary thermal degrees of freedom at chemical freeze-out \cite{ALICE2021, ALICE2023, ALICE2025}.
Instead, their production proceeds through a time-ordered nonequilibrium cascade mediated by short-lived intermediate resonances, most notably the $\Delta$.
A closely analogous pattern arises in cosmological Bose--Einstein condensation (BEC) of scalar-field dark matter, where coherent condensate cores form through repeated collapse and re-expansion events rather than simple equilibrium condensation \cite{FMT2008, FM2009}.

In previous work \cite{Fuku1}, these systems were successfully described using reduced rate equations for coarse-grained number densities.
While such equations capture the essential physics of delayed formation and survival, their microscopic status and domain of validity remain unclear.
Rate equations are intrinsically Markovian: the evolution at a given time depends only on the instantaneous state of the system.
However, intermediate reservoirs such as $\Delta$ resonances or collapse-induced localized structures have finite lifetimes and can temporarily store correlations.
This raises the question of when memory effects can be neglected, and what new phenomena arise beyond the Markovian approximation.

The purpose of this paper is to place the cascade picture on a systematic footing by embedding it within the Schwinger--Keldysh nonequilibrium field-theoretic framework \cite{Schwinger1961, Keldysh1965}.
We demonstrate that the familiar rate equations emerge as a controlled Markovian limit of the Keldysh description, clarify the origin of linear and quadratic transfer terms, and show how non-Markovian corrections encode the finite lifetime of intermediate reservoirs.
By comparing heavy-ion collisions and cosmological BEC within the same formalism, we identify a universal nonequilibrium structure governed by intermediate reservoirs and their associated memory effects.
The central message of this work is that the validity of commonly used
rate equations in cascade dynamics is controlled by the finite lifetime
of intermediate reservoirs, which sets a memory time governing when
non-Markovian effects become essential.

This paper is organized as follows.
In Sec.~II, we briefly review the Schwinger--Keldysh real-time formalism
and introduce the notation used throughout this work.
Sec.~III is devoted to the Markovian limit of the Kadanoff--Baym
equations and to the emergence of rate equations.
In Sec.~IV, we discuss an illustrative application of the general
framework to relativistic heavy-ion collisions.
Sec.~V constitutes the central part of this work, where we develop a
reduced non-Markovian cascade description relevant for nonequilibrium
condensation phenomena, with particular emphasis on the Bose--Einstein condensate cosmology, and identify the
memory time as a criterion for the validity of rate-based approaches.
Appendix~A provides an illustrative example that clarifies how a finite
reservoir lifetime induces non-Markovian effects and how the Markovian
limit is recovered.

\section{Schwinger--Keldysh framework}
\label{sec:SK}
The Schwinger--Keldysh, or closed-time-path (CTP), formalism provides a
natural language for describing nonequilibrium quantum field theory.
Expectation values are computed along a contour that evolves forward
and backward in real time, which leads to a doubling of degrees of
freedom.
The basic contour-ordered two-point function $G(x,y)$ can be reorganized
into retarded ($G^{R}$), advanced ($G^{A}$), and Keldysh ($G^{K}$)
components, which are particularly convenient for kinetic
interpretations.
For later use we recall the standard relations between the
contour-ordered Green functions and the retarded, advanced, and
Keldysh components obtained by the standard Keldysh rotation:
\begin{align}
G^{R} &= G^{--} - G^{-+}, \qquad
G^{A} = G^{--} - G^{+-}, \qquad
G^{K} = G^{-+} + G^{+-}.
\label{eq:ra_relations}
\end{align}
Here we employ the standard contour-ordered Green functions on the
closed-time path,
\begin{align}
G^{ab}(x,y) \equiv -i
\langle T_{c}\,\psi_{a}(x)\psi^{\dagger}_{b}(y)\rangle,
\qquad a,b=\pm ,
\end{align}
with $T_{c}$ denoting ordering along the closed-time contour.

With our choice of contour orientation, the diagonal components reduce
to
\begin{align}
G^{--}(x,y) &= -i \langle T\,\psi(x)\psi^{\dagger}(y)\rangle , \\
G^{++}(x,y) &= -i \langle \tilde T\,\psi(x)\psi^{\dagger}(y)\rangle ,
\end{align}
where $T$ and $\tilde T$ denote the usual time- and anti-time-ordering,
respectively.
The off-diagonal components define the lesser and greater Green
functions,
\begin{align}
G^{<}(x,y) \equiv G^{+-}(x,y)
 &= i \langle \psi^{\dagger}(y)\psi(x)\rangle , \\
G^{>}(x,y) \equiv G^{-+}(x,y)
 &= -i \langle \psi(x)\psi^{\dagger}(y)\rangle .
\end{align}
(This assignment follows our contour convention; physical results are
independent of this choice.)

The dynamics of the system is encoded in the Dyson--Schwinger equation,
\begin{equation}
G^{-1} = G_{0}^{-1} - \Sigma ,
\label{eq:Dyson}
\end{equation}
where $G_{0}$ denotes the free propagator and $\Sigma$ the self-energy.
In nonequilibrium situations both $G$ and $\Sigma$ are generally
nonlocal in time.
This nonlocality provides the microscopic origin of memory effects:
the evolution at time $t$ depends on the full past history of the system
through convolution integrals involving $\Sigma(t,t')$.

\section{From Keldysh equations to kinetic equations}

\subsection{Kadanoff--Baym equations and memory kernels}
The explicit rate equations emerge only after further coarse graining,
which will be carried out in Sec.~V; here we derive the non-Markovian
kinetic equations that form their microscopic starting point.
The equations of motion for the lesser and greater Green functions, $G^{<,>}$, are given by the Kadanoff--Baym equations \cite{KB1962}.
Schematically, for a homogeneous system they can be written as
\begin{equation}
\left(\partial_t + \ldots \right) G^{<}(t,t') =
\int \dd t_1 \, \Big[ \Sigma^{R}(t,t_1) G^{<}(t_1,t') + \Sigma^{<}(t,t_1) G^{A}(t_1,t') \Big],
\label{eq:KBschematic}
\end{equation}
with analogous expressions for $G^{>}$.
The explicit time integral demonstrates that the evolution depends on the full past history of the system through memory kernels.

\subsection{Markovian limit}
To obtain kinetic or rate equations, one typically performs a Wigner
transformation and a gradient expansion, followed by a quasiparticle
approximation.
If the correlation time associated with the self-energy is short compared
to the macroscopic evolution time, the memory kernel becomes sharply
peaked, reflecting a separation of time scales in the sense of
Zwanzig~\cite{Zwanzig1960,Zwanzig2001, MoriZwanzigReview}, and can be approximated by a delta
function in time,
\begin{equation}
\Sigma(t,t') \ \longrightarrow\ \Sigma(t)\,\delta(t-t')
\qquad \text{(Markovian limit)}.
\end{equation}


In this limit Eq.~(\ref{eq:KBschematic}) reduces to a local-in-time kinetic equation, and further coarse-graining over momenta yields rate equations for number densities.
Physically, the Markovian approximation is controlled by a separation of time scales:
\begin{equation}
\tau_{\rm corr} \ll \tau_{\rm macro},
\end{equation}
where $\tau_{\rm corr}$ is the microscopic correlation (or reservoir) time and $\tau_{\rm macro}$ is the characteristic time scale of expansion/cooling.
In this limit, phase information and temporal correlations encoded in
the memory kernel are discarded, leaving a closed description in terms
of instantaneous occupation numbers.
\subsection{Origin of linear and quadratic transfer terms}
Within this framework, the structure of the rate equations is
transparent.
Linear terms arise from one-body decay processes encoded in the
imaginary part of the retarded self-energy, while quadratic terms
originate from two-body scattering contributions to $\Sigma^{<,>}$.
As will become explicit in Eqs.~(17)--(19) and Eqs.~(26)--(27), the
appearance of linear transfer terms in heavy-ion collisions and
quadratic recondensation terms in cosmological BEC reflects the
dominance of different microscopic processes, rather than a fundamental
discrepancy between the systems.
This observation also clarifies how seemingly different reduced
descriptions can represent the same underlying nonequilibrium structure
after coarse-graining.

In the on-shell, gradient-expanded, and Markovian limit, the kinetic
equation reduces schematically to a local rate equation of the form
\begin{equation}
\dot n(t)=S_{\rm in}(t)-\Gamma_{\rm out}\,n(t),
\label{ndot}
\end{equation}
where $S_{\rm in}(t)$ denotes an effective source term generated by
microscopic processes.
Its concrete realization in cascade systems will be discussed in
Sec.~V.

\section{Heavy-ion collisions: $\Delta$-mediated nonequilibrium cascade}

The formation of light nuclei in relativistic heavy-ion collisions
provides a concrete realization of a nonequilibrium cascade mediated
by short-lived intermediate resonances \cite{ScheiblHeinz1999}.
Here we formulate this process at the level of coarse-grained rate
equations, emphasizing its role as a concrete example of a more general
non-Markovian cascade structure to be developed in Sec.~V.

\subsection{Minimal three-component description}

We consider three relevant degrees of freedom:
free nucleons $N=(p,n)$, $\Delta$ resonances, and deuterons $d$.
At the coarse-grained level, their number densities
$n_N(t)$, $n_\Delta(t)$, and $n_d(t)$ obey
\begin{align}
\dot n_\Delta &= -\Gamma_\Delta\, n_\Delta + S_\Delta(t),
\label{eq:ndelta_IV}
\\
\dot n_N &= \Gamma_\Delta\, n_\Delta
           - \lambda_{NN\to d}(T)\, n_N^2
           + S_N(t),
\label{eq:nN_IV}
\\
\dot n_d &= \lambda_{NN\to d}(T)\, n_N^2
           - \Gamma_d^{\rm br}(T)\, n_d .
\label{eq:nd_IV}
\end{align}
Here $\Gamma_\Delta$ denotes the in-medium decay width of the $\Delta$,
$\Gamma_d^{\rm br}(T)$ is the temperature-dependent deuteron breakup rate,
and $\lambda_{NN\to d}(T)=\langle\sigma_{NN\to d} v_{\rm rel}\rangle$
encodes the coalescence probability.
The source terms $S_{\Delta,N}(t)$ represent particle production at
hadronization and subsequent dilution due to expansion.

\subsection{Time ordering and delayed formation}

Equations~(\ref{eq:ndelta_IV})--(\ref{eq:nd_IV}) exhibit a characteristic
time ordering.
Immediately after hadronization, $S_\Delta(t)$ is large and the
temperature is high, so that $\Gamma_d^{\rm br}(T)$ strongly suppresses
any deuteron population.
As the system expands and cools, $S_\Delta(t)$ decreases, but a finite
population of $\Delta$ resonances remains.
Their decay injects correlated nucleons at later times,
when $\Gamma_d^{\rm br}(T)$ has already dropped.

This delayed injection is the key physical mechanism enabling deuteron
survival.
Formally, the solution of Eq.~(\ref{eq:ndelta_IV}) can be written as
\begin{equation}
n_\Delta(t)
= \int_{t_0}^{t} \dd t'\,
\ee^{-\Gamma_\Delta (t-t')}\, S_\Delta(t'),
\label{eq:ndelta_formal}
\end{equation}
which shows explicitly that the $\Delta$ population acts as a memory
kernel for baryon-number transfer. The finite lifetime of the $\Delta$ resonance therefore induces a
non-Markovian structure in the effective nucleon dynamics.
A minimal illustration of how this memory effect enters the
coalescence parameter $B_2$ is given in Appendix~A.

Substituting Eq.~(\ref{eq:ndelta_formal}) into Eq.~(\ref{eq:nN_IV})
yields an integro-differential equation for $n_N(t)$,
demonstrating that the reduced two-component description
$(n_N,n_d)$ is generically non-Markovian.

\subsection{Markovian limit}

If the $\Delta$ lifetime,
$\tau_\Delta \sim \Gamma_\Delta^{-1}$,
is short compared to the macroscopic expansion time scale,
the memory kernel in Eq.~(\ref{eq:ndelta_formal}) becomes sharply peaked.
In this limit, the nucleon source term becomes effectively local in time,
and Eqs.~(\ref{eq:nN_IV}) and (\ref{eq:nd_IV}) reduce to the familiar
Markovian rate equations used in coalescence models.

Thus, the commonly employed rate-equation description of light-nuclei
production corresponds to a controlled Markovian approximation.
Non-Markovian corrections encode the finite lifetime of the $\Delta$
resonance and may become relevant when $\tau_\Delta$ is comparable to
the expansion or cooling time scale.
From an experimental perspective, non-Markovian effects associated with the
finite $\Delta$ lifetime may manifest themselves as deviations from a smooth,
monotonic evolution of light-nuclei observables.
In particular, delayed or non-monotonic behavior of the coalescence parameter
$B_2$ as a function of collision centrality or
Here $B_2$ denotes the standard coalescence parameter,
defined through the relation
$E_d d^3N_d/dp_d^3 = B_2 (E_p d^3N_p/dp_p^3)(E_n d^3N_n/dp_n^3)$,
and measures the degree of phase-space correlation between nucleons
at the time of deuteron formation.
Here $B_2$ denotes the standard coalescence parameter,
defined through the relation
$E_d d^3N_d/dp_d^3 = B_2 (E_p d^3N_p/dp_p^3)(E_n d^3N_n/dp_n^3)$,
and measures the degree of phase-space correlation between nucleons
at the time of deuteron formation.

The coalescence parameter $B_2$ provides a direct observational handle
on the nonequilibrium cascade discussed above.
By definition, $B_2$ measures the degree of phase-space correlation
between nucleons at the time of deuteron formation.
In a purely Markovian description, $B_2$ is determined locally by the
instantaneous temperature and density of the hadronic medium,
$B_2(t) \simeq B_2\!\left(T(t)\right)$.

However, in the presence of an intermediate reservoir with a finite
lifetime, such as the $\Delta$ resonance, this locality assumption
breaks down.
As shown by Eq.~(\ref{eq:ndelta_formal}), the $\Delta$ population
stores baryon-number correlations over a time interval of order
$\tau_\Delta \sim \Gamma_\Delta^{-1}$.
As a consequence, the nucleon correlations relevant for deuteron
formation at time $t$ depend on the system history at earlier times
$t'<t$.

This implies that the coalescence parameter generically acquires a
non-Markovian structure,
\begin{equation}
B_2(t)
=
\int_{t_0}^{t} \dd t'\,
K_\Delta(t-t')\,
\mathcal{F}\!\left[T(t'),n_N(t')\right],
\label{eq:B2_memory}
\end{equation}
where $K_\Delta$ is a memory kernel controlled by the $\Delta$ lifetime.
In the limit $\tau_\Delta \ll \tau_{\rm macro}$,
$K_\Delta(t-t') \propto \delta(t-t')$, and the standard Markovian
coalescence picture is recovered.

Equation~(\ref{eq:B2_memory}) makes explicit that $B_2$ is not merely a
static parameter but a dynamical quantity encoding the memory of the
intermediate nonequilibrium reservoir.
This observation provides a direct link between experimental
light-nuclei observables and the non-Markovian cascade framework, which will be developed in Sec.~V.

\section{Our approach: reduced non-Markovian cascade description}

The purpose of this section is to clarify how a reduced,
non-Markovian cascade description emerges from a full
three-component nonequilibrium dynamics, and under what conditions it
further reduces to conventional rate equations.
Our approach is based on a systematic elimination of an intermediate
reservoir, while keeping track of the memory effects induced by its
finite lifetime.

A concrete illustration of how the finite lifetime of the intermediate
reservoir translates into an observable non-Markovian effect is
presented in Appendix~A, where the coalescence parameter $B_2$ is shown
to acquire a memory-kernel structure.
For heavy-ion applications, the finite lifetime of the intermediate
reservoir can be identified with the lifetime of the $\Delta$
resonance, $\tau\sim\Gamma_\Delta^{-1}$.
This provides a natural criterion for the validity of the Markovian
approximation in terms of the ratio $\tau/\tau_{\rm exp}$.

\subsection{Three-component formulation}

At the microscopic or mesoscopic level, both heavy-ion collisions
and cosmological Bose--Einstein condensation can be described by
three interacting components.
In the cosmological context \cite{FMT2008}, these are the boson gas
$\rho_g$, the coherent field $\rho_\phi$, and the localized component
$\rho_l$, whose evolution is governed by
\begin{align}
\dot{\rho}_g &= -3H\rho_g - \Gamma\,\rho_g, \label{eq:rho_g}\\
\dot{\rho}_\phi &= -6H(\rho_\phi - V)
+ \Gamma\,\rho_g - \Gamma'\,\rho_\phi, \label{eq:rho_phi}\\
\dot{\rho}_l &= -3H\rho_l + \Gamma'\,\rho_\phi. \label{eq:rho_l}
\end{align}
An analogous structure appears in heavy-ion collisions, with
$\rho_g \leftrightarrow n_N$, $\rho_\phi \leftrightarrow n_d$,
and $\rho_l \leftrightarrow n_\Delta$.

The defining feature of this formulation is the presence of an
intermediate component, $\rho_l$ (or $n_\Delta$), which temporarily
stores energy and correlations and mediates the transfer between
the initial and final states.
This intermediate reservoir introduces an intrinsic time delay
into the cascade dynamics.

\subsection{Integrating out the intermediate reservoir}

We now derive a reduced description by formally eliminating the
intermediate component.
For definiteness, consider Eq.~(\ref{eq:rho_l}), whose formal solution
is given by
\begin{equation}
\rho_l(t)
= \int_{t_0}^{t} \dd t'\,
\ee^{-3\int_{t'}^{t} H(\tau)\dd\tau}\,
\Gamma'\,\rho_\phi(t').
\label{eq:rhol_formal}
\end{equation}
Substituting this expression back into the equation for $\rho_\phi$
generates an integro-differential equation of the form
\begin{equation}
\dot{\rho}_\phi(t)
= \Gamma\,\rho_g(t)
- \int_{t_0}^{t} \dd t'\,
K(t,t')\,\rho_\phi(t'),
\label{eq:nonmarkov_phi}
\end{equation}
where $K(t,t')$ is a memory kernel determined by the lifetime and
dilution of the intermediate component.
Equation~(\ref{eq:nonmarkov_phi}) explicitly exhibits the
non-Markovian character of the reduced dynamics:
the evolution at time $t$ depends on the full past history of
$\rho_\phi$.

\subsection{Markovian limit and emergence of rate equations}

To obtain kinetic or rate equations, one typically performs a Wigner
transformation and a gradient expansion of the Kadanoff--Baym equations,
followed by a quasiparticle approximation.
If the correlation time associated with the self-energy is short
compared to the macroscopic evolution time, the memory kernel becomes
sharply peaked and can be approximated by a delta function in time,
\begin{equation}
\Sigma(t,t') \ \longrightarrow\ \Sigma(t)\,\delta(t-t')
\qquad \text{(Markovian limit)}.
\end{equation}
In this limit Eq.~(\ref{eq:KBschematic}) reduces to a local-in-time
kinetic equation.
Further coarse graining over momenta then yields rate equations for
number densities.

Physically, the validity of the Markovian approximation is controlled
by a separation of time scales,
\begin{equation}
\tau_{\rm corr} \ll \tau_{\rm macro},
\end{equation}
where $\tau_{\rm corr}$ denotes the microscopic correlation (or
reservoir) time encoded in the self-energy, and $\tau_{\rm macro}$ is
the characteristic time scale of expansion or cooling.
In this limit, phase information and temporal correlations contained in
the memory kernel are discarded, leaving a closed description in terms
of instantaneous occupation numbers.

At the level of the Kadanoff--Baym equations, the collision terms are
generically nonlinear.
For two-body scattering processes the lesser self-energy has the
schematic structure $\Sigma^{<}\propto G^{<}G^{<}$, while higher-order
processes may generate even higher powers of $G^{<}$.
When inserted into the Kadanoff--Baym equations, such contributions can
lead to collision terms that are cubic or higher order in $G^{<}$.

In the reduced description adopted here, we coarse-grain over momenta
and retain the dominant binary recondensation channel.
For heavy-ion applications, the intermediate-reservoir lifetime can be
identified with the lifetime of the $\Delta$ resonance,
$\tau\sim\Gamma_\Delta^{-1}$, which provides a natural criterion for the
validity of the Markovian approximation in terms of $\tau/\tau_{\rm exp}$.

After performing a Wigner transformation, an on-shell projection, and
integration over momenta, one of the Green functions entering the
collision integral is absorbed into an effective rate coefficient.
As a result, the net formation rate of the condensed component scales
quadratically with the density of the excited component.

At this level, the reduced dynamics can be written in terms of a local
Markovian rate equation of the form
\begin{equation}
\dot n_\phi
= S_{\rm in}(t) - \Gamma_{\rm out}\,n_\phi,
\label{eq:markov_nphi}
\end{equation}

where $S_{\rm in}(t)$ denotes an effective source term generated by
microscopic processes involving the remaining degrees of freedom, as
introduced in Sec.~III.
For clarity, dilution terms such as $-3H\rho_l$ are omitted here, as they
only modify the detailed shape of the memory kernel and do not affect
the reduced cascade structure.

Upon further coarse graining, the remaining component $\rho_g$ may be
viewed as an excited reservoir characterized by a coarse-grained number
density $n_{\rm ex}$.
At this stage, Eq.~(\ref{eq:markov_nphi}) corresponds to the schematic
Markovian rate equation~(\ref{ndot}) discussed earlier.

In the present cascade setting, the effective source term $S_{\rm in}$
originates dominantly from binary processes among the excited degrees of
freedom.
Within an on-shell and Markovian kinetic description, the corresponding
gain term can therefore be written as
\begin{equation}
S_{\rm in}(t)\;\equiv\;\kappa_{\rm in}\,n_{\rm ex}^2(t),
\end{equation}
where $\kappa_{\rm in}$ is an effective two-body recondensation rate
coefficient that encodes the underlying microscopic dynamics after
integrating out the intermediate reservoir.

This identification leads to the minimal two-component description
\begin{align}
\dot n_0 &= \kappa_{\rm in}\, n_{\rm ex}^2
           - \Gamma_{\rm out}\, n_0, \label{eq:n0_final}\\
\dot n_{\rm ex} &= -\kappa_{\rm in}\, n_{\rm ex}^2
                  + \Gamma_{\rm out}\, n_0
                  + S_{\rm ex}(t), \label{eq:nex_final}
\end{align}

While the heavy-ion case involves a well-defined reaction rate
$\lambda_{NN\to d}$ associated with nucleon coalescence (Eqs.~(\ref{eq:nN_IV}) and (\ref{eq:nd_IV})), the
corresponding quantity in the Bose--Einstein condensation context is an
effective two-body recondensation coefficient, $\kappa_{\rm in}$,
which encodes binary scattering processes among excited modes.

A simple illustrative example of how a finite reservoir lifetime
induces non-Markovian effects, and how the Markovian limit is recovered,
is discussed in Appendix~A.

\subsection{Physical interpretation}

Equations~(\ref{eq:n0_final}) and (\ref{eq:nex_final}) are not introduced
as independent phenomenological assumptions.
They emerge as the Markovian and coarse-grained limit of the underlying
three-component nonequilibrium dynamics once the intermediate reservoir
has been integrated out.
The apparent change from linear transfer terms in
Eqs.~(\ref{eq:rho_g})--(\ref{eq:rho_l})
to quadratic terms in the reduced description directly reflects this
elimination procedure.
Linear terms describe one-body decay or transfer processes involving
the intermediate reservoir, whereas quadratic terms arise once this
reservoir is eliminated and the dynamics is governed by two-body
processes among the remaining degrees of freedom.
A characteristic observable consequence of the non-Markovian dynamics
is a delay in the formation of the final bound or coherent state.
Because the transfer of population proceeds through an intermediate
reservoir with a finite lifetime, the buildup of the final state does
not start immediately even when the microscopic production channels
are open.
Such a formation delay is absent in purely Markovian rate equations,
where the evolution is local in time and responds instantaneously to
changes in the source terms.
In practical applications, this delay can manifest itself as a
suppression or modification of coalescence observables, such as the
parameter $B_2$, compared to expectations based on instantaneous rate
equations.

Within this formulation, the distinction between the $\Delta$ resonance
in heavy-ion collisions and the localized component $\rho_l$ in the
cosmological BEC scenario becomes immaterial at the structural level.
Both act as intermediate nonequilibrium reservoirs characterized by a
finite memory time that delays the transfer of correlations from the
initial to the final state.
While the $\Delta$ resonance is unstable in the spectral sense, the
instability of $\rho_l$ is dynamical rather than associated with a decay
width.
Nevertheless, both play the same essential role: they temporarily store
energy and correlations and release them only after the environment has
evolved to a regime in which the final fragile state can survive.

From this perspective, the survival of weakly bound or coherent
structures is governed not by their binding energy alone, but by the
existence and lifetime of intermediate reservoirs.
The reduced rate equations capture this mechanism in the Markovian
limit, while the full non-Markovian description encodes it through
explicit memory kernels or, equivalently, through additional dynamical
degrees of freedom.
An explicit minimal model demonstrating how the finite lifetime of the
intermediate reservoir translates into a non-Markovian observable is
presented in Appendix~A, where the coalescence parameter $B_2$ is shown
to acquire a memory-kernel structure.
\section{Conclusions and outlook}

In this work we have analyzed nonequilibrium cascade dynamics in which
fragile bound or coherent structures are formed through intermediate
states, rather than by direct equilibration.
Using the Schwinger--Keldysh real-time formalism, we have shown how
commonly employed rate equations arise as a controlled Markovian
approximation of an underlying multi-component nonequilibrium dynamics.

Our central result is that the validity of such rate-based descriptions
is governed by the finite lifetime of intermediate reservoirs.
When this lifetime is short compared to macroscopic evolution scales,
the reduced dynamics becomes effectively Markovian and can be captured
by local rate equations.
When the lifetime is retained, however, non-Markovian memory effects
naturally emerge, leading to delayed and history-dependent formation
dynamics that cannot be described by instantaneous rates alone.
The associated memory time therefore provides a quantitative criterion
for assessing when reduced rate equations are reliable.

We have illustrated this general mechanism by connecting heavy-ion
light-nuclei production mediated by $\Delta$ resonances with
cosmological Bose--Einstein condensation involving localized
intermediate structures.
Despite their very different physical settings, both systems share the
same structural feature: an intermediate nonequilibrium reservoir that
temporarily stores energy and correlations and delays their transfer to
the final fragile state.
This common structure explains why apparent thermal or
equilibrium-like final yields can emerge even in the absence of genuine
equilibration.

From a broader perspective, the comparison between heavy-ion collisions
and cosmological Bose--Einstein condensation reveals a universal
nonequilibrium structure.
In both cases, intermediate reservoirs act as information bottlenecks
that temporarily store correlations and delay the formation of fragile
final states.
The commonly used Markovian rate equations correspond to the limit in
which these bottlenecks are effectively instantaneous, while the full
Schwinger--Keldysh description captures their finite duration and the
associated memory effects.
From this viewpoint, apparently thermal or equilibrium-like final
yields need not imply genuine equilibration.
Rather, they can arise as dynamical attractors of nonequilibrium
evolution once intermediate reservoirs are integrated out.
Non-Markovian effects then quantify the extent to which this
integration is incomplete and provide a systematic way to assess the
reliability of reduced rate descriptions.

From a phenomenological perspective, non-Markovian effects imply
observable signatures beyond those captured by conventional rate
equations.
In particular, the finite lifetime of the intermediate reservoir leads
to a delayed buildup of the final state, an effect that is absent in
purely Markovian descriptions.
Such delays can modify coalescence observables, as explicitly
demonstrated in Appendix~A for the parameter $B_2$, and provide a
natural target for future experimental and phenomenological studies.

Looking ahead, the reduced non-Markovian cascade framework developed
here offers a systematic way to assess the applicability of rate
equations in a wide range of nonequilibrium processes.
While our discussion has focused on heavy-ion collisions and
cosmological condensation, the same reasoning applies to other systems
in which intermediate states play a central dynamical role.
Extending this approach to more detailed microscopic models and to
quantitative comparisons with experimental data constitutes an
important direction for future work.
\medskip

\noindent
{\bf Acknowledgments}  
This work is supported in part by 
  Grant-in-Aid for Science Research from the Ministry of Education, Science and Culture, Japan (No.~25H00653).
\appendix
\section{Minimal model for the memory dependence of the coalescence parameter $B_2$}

In this Appendix we illustrate, within a minimal model, how the finite
lifetime of the intermediate $\Delta$ resonance induces a non-Markovian
structure in the coalescence parameter $B_2$.

\subsection{Memory kernel from $\Delta$ decay}

As discussed in Sec.~IV, the $\Delta$ population acts as an intermediate
reservoir that stores baryon-number correlations over a characteristic
time scale $\tau_\Delta \sim \Gamma_\Delta^{-1}$.
The formal solution of the $\Delta$ rate equation,
Eq.~(\ref{eq:ndelta_IV}), leads to an exponential memory kernel,
\begin{equation}
K_\Delta(t-t')
=
\Gamma_\Delta\,\ee^{-\Gamma_\Delta (t-t')}\,\theta(t-t').
\label{eq:Kdelta}
\end{equation}
This kernel describes the delayed release of correlated nucleons
originating from $\Delta$ decay.

\subsection{Effective expression for $B_2(t)$}

The coalescence parameter $B_2$ measures the probability that two nucleons
with appropriate quantum numbers and relative momentum form a deuteron.
In the presence of a finite memory time, this probability depends on the
history of nucleon injection.
A minimal phenomenological expression consistent with the discussion in
Sec.~IV is
\begin{equation}
B_2(t)
=
\int_{t_0}^{t} \dd t'\,
K_\Delta(t-t')\,
\mathcal{P}_{NN\to d}\!\left[T(t')\right],
\label{eq:B2_minimal}
\end{equation}
where $\mathcal{P}_{NN\to d}(T)$ denotes the instantaneous probability for
two nucleons to coalesce into a deuteron at temperature $T$.

\subsection{Markovian limit}

In the limit $\Gamma_\Delta \to \infty$,
the kernel~(\ref{eq:Kdelta}) approaches a delta function,
\begin{equation}
K_\Delta(t-t') \;\longrightarrow\; \delta(t-t'),
\end{equation}
and Eq.~(\ref{eq:B2_minimal}) reduces to the familiar Markovian result
\begin{equation}
B_2(t) \simeq \mathcal{P}_{NN\to d}\!\left[T(t)\right].
\end{equation}
This corresponds to the standard coalescence picture in which $B_2$ is
determined solely by the instantaneous properties of the hadronic medium.

\subsection{Non-Markovian features}

For finite $\Gamma_\Delta$, Eq.~(\ref{eq:B2_minimal}) predicts several
generic non-Markovian features.
First, the effective formation of deuterons is delayed relative to the
cooling of the medium.
Second, if $\mathcal{P}_{NN\to d}(T)$ varies rapidly with time,
the convolution with $K_\Delta$ can lead to a non-monotonic behavior of
$B_2(t)$.
Both effects directly reflect the finite memory time associated with the
intermediate $\Delta$ reservoir.

These features provide a clear criterion for identifying non-Markovian
effects in experimental data:
deviations from a smooth, monotonic dependence of $B_2$ on collision
centrality or system size signal the breakdown of the instantaneous
coalescence approximation.


\end{document}